\documentclass[aps,longbibliography,showpacs,twocolumn,superscriptaddress,amsmath,amssymb,verbatim]{revtex4-1}

\usepackage[thinlines]{easytable}
\usepackage{graphicx}
\usepackage{lmodern}
\usepackage{epstopdf}
\usepackage{dcolumn}
\usepackage{bm}
\usepackage{subfigure}
\usepackage{makecell}
\usepackage{amsmath} 
\usepackage[pdftex,colorlinks=true,citecolor=blue,linkcolor=blue,urlcolor=blue,bookmarks=true]{hyperref}

\usepackage{color}
\usepackage{textcomp}
\definecolor{red}{rgb}{1,0,0}

\definecolor{blue}{rgb}{0,0,1}

\definecolor{green}{rgb}{0,1,0}

\begin{document}
	\preprint{APS}

	\title{Spin liquid state in a rare-earth hyperkagome lattice }
\author{J. Khatua}
\affiliation{Department of Physics, Indian Institute of Technology Madras, Chennai 600036, India}
\author{S. Bhattacharya} 
\affiliation{Université Paris-Saclay, CNRS, Laboratoire de Physique des Solides, 91405, Orsay, France}

\author{Q. P. Ding } 
\affiliation{Ames Laboratory, U.S. DOE, and Department of Physics and Astronomy, Iowa State University, Ames, Iowa 50011, USA}
\author{S. Vrtnik}
\affiliation {Jo\v{z}ef Stefan Institute, Jamova c. 39, 1000 Ljubljana, Slovenia}
	\author{A. M. Strydom}
\affiliation{Highly Correlated Matter Research Group, Department of Physics, University of Johannesburg, PO Box 524, Auckland Park 2006, South Africa}
\author{N. P. Butch}
\affiliation{NIST Centre for Neutron Research, Gaithersburg, Maryland, USA}
\author{H. Luetkens}
\affiliation{Laboratory for Muon-Spin Spectroscopy, Paul Scherrer Institute, CH-5232 Villigen-PSI, Switzerland}
\author{E. Kermarrec}
\affiliation {Université Paris-Saclay, CNRS, Laboratoire de Physique des Solides, 91405, Orsay, France}
\author{M. S. Ramachandra Rao}
\affiliation{Department of Physics, Nano Functional Materials Technology Centre and Materials Science Research Centre, Indian Institute of Technology Madras, Chennai-600036, India}
\affiliation{Quantum Centre for Diamond and Emergent Materials, Indian Institute of Technology Madras,
	Chennai 600036, India.}

\author{ A. Zorko}
\affiliation {Jo\v{z}ef Stefan Institute, Jamova c. 39, 1000 Ljubljana, Slovenia}
\affiliation{Faculty of Mathematics and Physics, University of Ljubljana, Jadranska u. 19, 1000 Ljubljana, Slovenia}
\author{ Y. Furukawa}
\affiliation{Ames Laboratory, U.S. DOE, and Department of Physics and Astronomy, Iowa State University, Ames, Iowa 50011, USA}
\author{P.Khuntia}
\email[]{pkhuntia@iitm.ac.in}
\affiliation{Department of Physics, Indian Institute of Technology Madras, Chennai 600036, India}
\affiliation{Quantum Centre for Diamond and Emergent Materials, Indian Institute of Technology Madras,
	Chennai 600036, India.}
\affiliation{Functional Oxide Research Group, Indian Institute of Technology Madras, Chennai 600036,
	India.}
\date{\today}
\begin{abstract}
		Quantum fluctuations enhanced by frustration and subtle interplay between competing degrees of freedom offer an ideal ground to realize novel states with fractional quantum numbers in quantum materials that defy standard theoretical paradigms. Quantum spin liquid (QSL) is a highly entangled state wherein frustration induced strong quantum fluctuations preclude symmetry breaking phase transitions down to zero temperature without any order parameter.  Experimental realizations of QSL in quantum materials with spin dimensionality greater than one is very rare. Here, we present our  thermodynamic, nuclear magnetic resonance,  muon spin relaxation and inelastic neutron scattering  studies of a new rare-earth hyperkagome  compound Li$_{3}$Yb$_{3}$Te$_{2}$O$_{12}$ in which Yb$^{3+}$ ions constitute a three dimensional spin-lattice without any detectable disorder. Our comprehensive experiments evince neither signature   of  magnetic ordering nor spin freezing down to 38 mK that suggest the realization of  dynamic liquid-like ground  state  in this antiferromagnet. The  ground state of this material is interpreted by a low energy $J_{\rm eff}$ = 1/2  degrees of freedom with short range spin correlations.	  
	The present results demonstrate a viable basis to explore spin-orbit driven enigmatic correlated  quantum states in a new class of rare-earth based three dimensional frustrated magnets that may open  new avenues in theoretical and experimental  search for spin liquids.
\end{abstract}
\maketitle
\section{Introduction}
Quantum materials wherein superposition and entanglement are at play may exhibit exotic physical  phenomena, such as spin liquids with fractional quantum numbers coupled to emergent gauge field that offer a novel paradigm  in advancing quantum science and technology \cite{ref1,Tokura2017,Basov2017,Keimer2017}.
The experimental realization of quantum spin liquid with exotic collective excitations in quantum  materials set an outstanding track in modern condensed matter since the seminal proposal of P. W. Anderson in 1973 \cite{ANDERSON1973153,Balents2010}.
Quantum spin liquids (QSLs) are  characterized
by absence of magnetic order down to  $T = 0 $, long-range entanglement,  coherent fluctuation of spins and  the preservation of local symmetries despite strong exchange interaction between spins \cite{Balents2010,Gingras_2014,Takagi2019d,RevModPhys.88.041002}. 
Fractionalization of quantum numbers is a fingerprint of QSLs which are different from spin waves in conventional magnets. This is a well established scenario in one-dimensional (1D) magnets. It has been suggested that also 2D and 3D QSLs can host fractionalized deconfined quasiparticles such as spinons  and Majorana fermions the understanding of which is essential in advancing fundamental physics and is highly relevant for fault-tolerant quantum computing \cite{Balents2010,Nayak_2008,Savary_2016,Broholmeaay0668,RevModPhys.89.025003,klanjvsek2017high,PhysRevLett.117.097201,Khuntia2020,Han2012}.    
However, the presence of  perturbations, such as extra exchange couplings, unavoidable disorder,
and defects in real materials put a strong constraint for the ideal realization of quantum spin liquids  with D $\geq$ 2 \cite{PhysRevLett.69.2590,PhysRevLett.119.157201,PhysRevB.93.140408}.\\
Moreover,  3D spin lattices that are composed of small triangular motifs in edge- or corner-sharing fashion, can host strong frustration similar to 2D spin lattices.
Namely, quantum fluctuations in such 3D frustrated lattices with low  value of spin could lead to unconventional ground states including spin liquids \cite{Savary_2016,Broholmeaay0668}.
This has  motivated us  to explore new quantum materials  with rich potential to host such states with  fractional quantum numbers \cite{Knolle2019,PhysRevLett.116.107203,Chillal2020}.
The intertwining of spin-orbit coupling, anisotropy, spin correlations and  frustration in rare-earth magnets offer another route to realize novel quantum states \cite{Arh2022,Gingras_2014,Keimer2017,annurev-conmatphys-020911-125138,PhysRevX.9.021017}. 
In recent years,  spin-orbit driven frustrated magnets  have provided a new light to realize novel quantum phenomena  ranging from spin ice  state \cite{Ramirez1999,Gingras_2014} to spin liquid \cite{Arh2022,PhysRevLett.117.097201,Bordelon2019} and  anomalous and spontaneous  Hall effect \cite{Taguchi2573,Machida2010} to Bose-Einstein
condensate phase \cite{PhysRevLett.123.027201,annurev-conmatphys-020911-125138}.  In  lanthanide magnetic materials, the interplay between spin-orbit coupling and low-symmetry crystal electric field  leads to Kramers  doublets  where effective $J_{\textnormal{eff}}$ = 1/2 moments of rare-earth ions appear, contrary to  pure $S$ = 1/2 moments in transition metal ions. The 3D spin lattice based quantum magnets host a plethora of  unconventional properties of 3D magnets including non-trivial short-range spin correlation \cite{PhysRevLett.91.167201}, unconventional spin glass \cite{PhysRevB.89.054433}, and magnetic Coulomb phase \cite{Fennell415,Petit2016}. 
For example, the lanthanide family A$_{2}$B$_{2}$O$_{7}$ 
where rare-earth ions form a pyrochlore lattice is a  rich reservoir that hosts a wide range of quantum phenomena owing to complex interplay between competing degrees of freedom \cite{Gingras_2014,Bramwell1495}.  
Quite recently, Ce$_{2}$Zr$_{2}$O$_{7}$ was proposed to be a 3D QSL candidate, where Ce$^{3+}$ ions with $J_{\textnormal{eff}}$ = 1/2 moments decorate a
network of corner-sharing tetrahedra \cite{Gao2019}. The observation of $U$(1)  QSL  and emergent phases driven by  multipolar interactions in its analogue  Ce$_{2}$Sn$_{2}$O$_{7}$ also set a unique example of a 3D spin-lattice for exploring quantum many-body phenomena \cite{Sibille2020}.
Furthermore, the garnet series R$_{3}$B$_{5}$O$_{12}$ (R = rare-earth ions and B = Ga and Al) where the rare-earth ions decorate two interpenetrating hyperkagome lattices are  known to harbor myriads of interesting quantum phenomena such as long-range multipolar state \cite{Paddison179,PhysRevB.104.064425} and Ising antiferromagnets \cite{PhysRevB.100.184415,PhysRevB.105.014441}. However,  most of the garnets undergo a magnetic phase transition at lower temperature \cite{PhysRev.156.663,PhysRevB.91.014419,PhysRevB.100.094442,PhysRevB.39.11413,PhysRevB.93.094419,PhysRevB.69.064404,PhysRevB.100.241111}.   
Therefore, the search for an ideal 3D spin lattice to realize a QSL state poses a particularly challenging problem.
For example no report has been published so-far on a structurally perfect 4$f$-based hyperkagome lattice in which high degree of spin frustration and anisotropic interactions mediated by spin-orbit coupling could stabilize an enigmatic 3D spin liquid.\\  Here, we report a new rare-earth based quantum magnet Li$_{3}$Yb$_{3}$Te$_{2}$O$_{12}$ and  investigate its inherent physics of 3D frustrated spin-lattice with $J_{\textnormal{eff}}$ = 1/2 moments  of Yb$^{3+}$ ions. 
Li$_{3}$Yb$_{3}$Te$_{2}$O$_{12}$ crystallizes in the cubic space group $Ia\bar{3}d$ where Yb$^{3+}$ ions form  an interpenetrating hyperkagome spin-lattice. 
Our  thermodynamic results suggests the presence of a Kramers doublet state  and antiferromagnetic interaction  between $J_{\rm eff}$ = 1/2 spins at low temperature. The absence of long-range magnetic ordering and spin freezing  down to at least 38 mK in our ac susceptibility, specific heat and NMR experiments is the first fingerprint of a spin liquid state in this frustrated quantum material. Furthermore, specific heat measurements show the presence of short-range spin correlations in this antiferromagnet.
Our complementary experimental probes thus demonstrate a dynamic ground state and  a field-induced gapped behavior in  the spin excitation spectrum in this spin-orbit driven frustrated hyperkagome quantum material.
\begin{figure*}[!htbp]
	\centering
	\includegraphics[width=\textwidth]{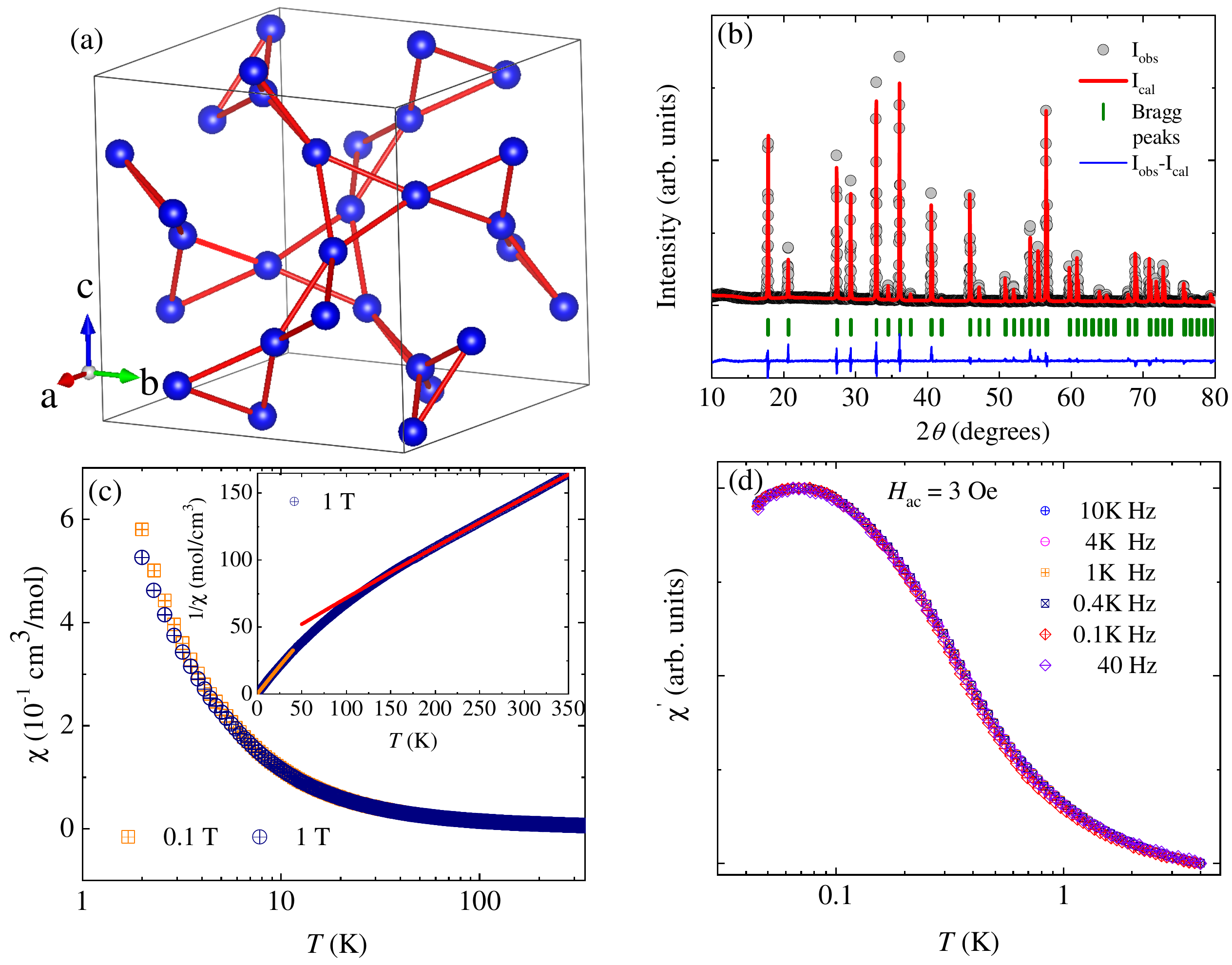}
	\caption{\label{st}Crystal structure and bulk magnetic properties of LYTO. (a) The position of the Yb$^{3+}$ magnetic ions  within the unit cell. These Yb$^{3+}$ ions decorate
		two interpenetrating chains of 3D corner-shared triangles known as the hyperkagome
		lattice. (b) Rietveld refinement of powder X-ray diffraction data recorded at room temperature. The experimentally observed points (circles), the calculated Rietveld refinement profile (solid line), the Bragg reflection positions (olive vertical bars) and the difference between the observed and calculated intensity (blue line) are shown. (c) The temperature dependence of  magnetic susceptibility in  two magnetic fields. The inset shows the temperature dependent inverse
		magnetic susceptibility data with the orange and red lines showing the Curie-Weiss fits for the
		low- and high-temperature data, respectively. (d) The temperature dependence
		of the real part of ac susceptibility at different frequencies down to 45 mK.  }
\end{figure*} 
\section{METHODS}
\textit{Sample synthesis:} Polycrystalline samples of Li$_{3}$Yb$_{3}$Te$_{2}$O$_{12}$ (henceforth LYTO) were synthesized by a standard solid-state reaction of Li$_{2}$CO$_{3}$ (99.0 \%, Alfa Aesar),  Yb$_{2}$O$_{3}$ (99.998 \%, Alfa Aesar), and  TeO$_{2}$ (99.9995 \%, Alfa Aesar) \cite{Cevallos_2018}.
Since the lanthanide-oxide and Li$_{2}$CO$_{3}$ are generally hygroscopic, before being weighted, Yb$_{2}$O$_{3}$ and Li$_{2}$CO$_{3}$ were preheated at 900$^\circ$C and 100$^\circ$C, respectively. The stoichiometric mixtures of starting materials were pelletised and heated at 800$^\circ$C for 30 hours in air
with several intermittent grindings. We also used Y$_{2}$O$_{3}$ (99.999 \%, Alfa Aesar) and followed the same procedure to prepare the isostructural non-magnetic analogue Li$_{3}$Y$_{3}$Te$_{2}$O$_{12}$.  The phase purity was checked by room temperature
powder X-ray diffraction (XRD) using smartLAB Rigaku  X-ray diffractometer with Cu K$\alpha$ radiation ($\lambda$ =1.54 {\AA}). The  crystal structure of LYTO was confirmed by the  Rietveld refinement of X-ray diffraction data using GSAS software \cite{Toby:hw0089}. All  the XRD peaks could be indexed with the cubic space group $Ia\bar{3}d$ and lattice parameter $a$ = 12.173 {\AA} \cite{Cevallos_2018}. No secondary phase is detected in XRD and the Rietveld refinement of XRD data rules out the presence of any  inter-site disorder between  the constituent atoms. The observed sharp XRD peaks indicate high quality polycrystalline samples investigated in this work.\\ 
\textit{Magnetization measurements:} Magnetization data  were acquired using a Quantum Design, SQUID magnetometer in the temperature range 2 K $\leq$ \textit{T} $\leq$ 350 K. Magnetic susceptibility ($\chi(\textit{T})$) data were recorded in magnetic fields 0.1 and 1 T and  the zero-field cooled (ZFC) and field cooled (FC) data were collected in a  magnetic field of 0.01 T. In order to determine the effective magnetic moment ($\mu_{\textnormal{eff}}$) and Curie-Weiss temperature ($\theta_{\textnormal{CW}}$), the high-temperature ($>$160 K) inverse susceptibility (1/$\chi(\textit{T})$) data were fitted with Curie-Weiss (CW) law 
$\chi$ = $\chi_{\textnormal{0}}$+ \textit{C}/(\textit{T} $-$ $\theta_{\textnormal{CW}}$). Here  \textit{C} is the Curie constant which is used to calculate $\mu_{\textnormal{eff}}=\sqrt{8C}$  $\mu_{\textnormal{B}}$, $\chi_{\textnormal{0}}$ is the temperature independent paramagnetic susceptibility, and $\theta_{\textnormal{CW}}$ provides an estimate of  magnetic exchange interactions. The isotherm  magnetization data are reproduced well (see SI Fig.~3) by a function of the form $M$/$M_{\textnormal{s}}$ = $B_{1/2}$ ($y$),  where $B_{J}(y) = [\frac{2J+1}{2J} coth[\frac{y(2J+1)}{2J}]-\frac{1}{2J}coth\frac{y}{2J}]$ is the Brillouin function, $M_{s}$ (= g$J\mu_{\textnormal{B}}$) is the saturation magnetization and the parameter $y=g\mu_{\textnormal{B}}J \mu_{\rm 0}H/k_{\rm B}T$, where $\mu_{\textnormal{B}}$ is the Bohr magneton, $g$ is the Land\'e  g-factor. The Brillouin function fit with fixed $J$ = 1/2 to magnetization data
yields Land\'e $g$ factor, $g$ = 3.54 $\pm$ 0.02.
\\
\textit{Specific heat measurements}: Specific heat measurements were performed in the temperature range 0.35 K $\leq$ \textit{T} $\leq$ 270 K under
magnetic fields 0 T $\leq$ $\mu_{\textnormal{0}}H$ $\leq$ 7 T, using thermal-relaxation technique provided by Quantum Design, PPMS. Furthermore, specific heat measurements
were carried out separately in the temperature range 0.054 K $\leq$ \textit{T} $\leq$ 4 K in zero-field  using a dilution refrigerator which was also used to measure ac susceptibility in the
temperature range 0.045 K $\leq$ $T$ $\leq$ 4 K at six different frequencies using a Dynacool PPMS
instrument from Quantum Design. The total specific heat of LYTO can be expressed as a sum of contributions from the electronic  spins  in the ground state Kramers  doublet ($C_{\rm mag}$), lattice contribution ($C_{\rm lat}$) and nuclear contribution ($C_{n}$) i.e., $C_{\textnormal{tot}}(T)=C_{\textnormal{mag}}(T) + C_{\textnormal{lat}}(T)$ + $C_{n}$.  We also measured specific heat of the Li$_{3}$Y$_{3}$Te$_{2}$O$_{12}$ which we used as a non-magnetic analogue to subtract the lattice contribution. After subtraction lattice contribution, we  subtracted nuclear Schottky contribution $C_{n}$ $\propto$ $T^{-2}$ to obtain magnetic specific heat.    \\
\textit{NMR measurements:} Field-swept $^{7}$Li ($I$ = 3/2, and gyromagnetic ratio 16.54 MHz/T) NMR measurements  down to 38 mK at several frequencies were carried out on a homemade phase-coherent spin-echo pulse spectrometer equipped with a 9 T Oxford magnet. The low  temperature NMR measurements were performed with a Oxford Kelvinox dilution refrigerator. NMR spectra measurements were carried out using a standard Han-echo sequence while the spin lattice relaxation time measurements were performed following saturation-recovery method. In LYTO, $^{7}$Li is an NMR active nucleus which couples with  Yb$^{3+}$ ions via hyperfine interactions. Therefore, $^{7}$Li NMR can  probe the intrinsic magnetic susceptibility as well as  low-energy spin excitations via  spectrum position, and spin-lattice relaxation, respectively.  The temperature dependence of the NMR shift ($K$)  at different frequencies was extracted from the fit of  field-swept NMR spectra. For an anisotropic magnetic material three different components of  NMR shift \textit{K}(\textit{T}) such as $K_{\textnormal{iso}}$ (isotropic shift), $K_{\textnormal{ax}}$ (axial shift) and  $K_{\textnormal{aniso}}$ (anisotropic shift), arising from a general hyperfine coupling tensor, can be determined from the observed experimental  line shape. In this respect, the principal components  $K_{\textnormal{X}}$, $K_{\textnormal{Y}}$ and $K_{\textnormal{Z}}$ of  tensor \textit{K}(\textit{T}) can be used to define 
$K_{\textnormal{iso}}$ = 1/3($K_{\textnormal{X}}$ $+$ $K_{\textnormal{Y}}$+K$_{\textnormal{Z}}$), $K_{\textnormal{aniso}}$ = 1/2($K_{\textnormal{Y}}$ $-$ $K_{\textnormal{X}}$) and \textit{K}$_{ax}$ = 1/6(2$K_{\textnormal{Z}}$ $-$ $K_{\textnormal{X}}$ $-$ $K_{\textnormal{Y}}$) \cite{Shimizu2012,PhysRevB.102.064429}.
The  temperature dependence of the NMR shift  scales with the magnetic susceptibility as
$	\textit{K}_{\textnormal{ax/iso}}(\textit{T})=\textit{K}_{0}+(A_{\textnormal{hf}}^{\rm ax/iso}/{N_{\textnormal{A}}\mu_{\textnormal{B}}})\chi_{\textnormal{spin}}(\textit{T})$,
where $K_{0}$ is the temperature independent chemical and orbital shift, $A_{\textnormal{hf}}$ is the hyperfine coupling constant between $^{7}$Li nucleus and Yb$^{3+}$ spins, $\mu_{\textnormal{B}}$ is the Bohr magneton and $N_{\textnormal{A}}$ is the Avogadro number, respectively. To extract the hyperfine couplings, the standard Clogston-Jaccarino plot (see SI Fig.~7)  was used to fit in two temperature  regimes and the obtained fitting parameters are given in SI note 5. 
   The fit of NMR spectra below 1 K is not very good (see SI Fig.~6) which could be due  to an additional spectral weight in  the low-field side  at low temperature,however, the origin is not clear at this moment. 
         To uncover the dynamics of Yb$^{3+}$ spins in LYTO at very low-energy, we conducted spin-lattice relaxation time measurement.
In order to estimate $T_{1}^{-1}$,
the recovery of longitudinal nuclear magnetization  data were fitted by the single exponential function  $	M_{z}(t)=(M_{0}-M(t))/M_{0}= A \ \ \textnormal{exp}(-t/{\textit{T}_{1}})$
in the entire temperature range of the investigation, where $M_{\textnormal{0}}$ is the equilibrium magnetization, $M_{z}(t)$ is the magnetization at time \textit{t} after the saturation pulse and $A$ is a constant.  The validity of a single exponential function in a wide temperature range suggests that the electronic moments are uniformly distributed in the host lattice without disorder. The fits of the  recovery curves with a formula for $I$ = 3/2 nuclei were  not as good as the single exponential function.
\\
\textit{$\mu$SR:} Muon spin relaxation measurements were performed at the Paul Scherrer Institute, on the GPS instrument. A weak transverse field of 50~Oe was applied to determine the initial asymmetry parameter. Zero-field (ZF) and longitudinal applied fields (LF) datasets were obtained using VETO mode resulting in minimal background signal ($<0.02$).\\
\textit{Inelastic neutron scattering:} Time-of-flight inelastic neutron scattering measurements were carried out at the NIST Center for Neutron Research (NCNR, Gaithersburg, MD) on the DCS spectrometer. In order to isolate the field evolution of the inelastic spectrum, a field-independent background, made of the average of the 0 and 2T datasets for $E > 0.85$~meV and of the average of the 7, 8, 9 and 10 T datasets for  $E < 0.85$~meV, was subtracted from the $Q$-integrated energy cuts.  Simulations of powder-averaged dynamical structure factor for inelastic neutron scattering were done with the SpinW software \cite{Toth_2015}. 
\section{Results}
\subsection{Magnetization}
The magnetic susceptibility $\chi$(\textit{T}) data do not exhibit any signs of a magnetic phase transition down to 2 K (Fig.~\ref{st} (c)).  The Curie-Weiss (CW) fit to the high-temperature ($>$160 K) $\chi(T)$ data yields $\theta_{\textnormal{CW}}$ $\approx$ $-$78 (1) K, $\mu_{\textnormal{eff}}$ = 4.4 $\mu_{\textnormal{B}}$, which is close to $\mu_{\textnormal{eff}}$ = 4.54 $\mu_{\textnormal{B}}$ of a 
free  Yb$^{ 3+}$ (\textit{J} = 7/2) spin and $\chi_{\rm 0}$ = 4.97 $\times$ 10$^{-4}$ cm$^{3}$/mol. The estimated large negative $\theta_{\textnormal{CW}}$ is ascribed  to the energy scale of excited crystal electric field levels \cite{Arh2022}. The interaction energy scale of 4\textit{f} systems is much lower and is in our case revealed at 
low-temperature where the CW fit of 1/$\chi$(\textit{T}) data in the temperature range 2 K $\leq$ \textit{T} $\leq$ 10 K  yields $\mu_{\textnormal{eff}}$ = 3.0 $\mu_{\textnormal{B}}$ and $\theta_{\textnormal{CW}}$ $\approx$ $-$0.3 K $\pm$ 0.03 K.  The obtained effective moment is smaller than the Yb$^{3+}$ free-ion value, which suggests a Kramers  doublet state  at low temperature, as also confirmed by specific heat measurements (see below). The negative CW temperature indicates  antiferromagnetic interactions  between the pseudo 1/2 moment of Yb$^{3+}$ ions with approximate interaction energy of the order of $\theta_{\textnormal{CW}}$. The Curie-Weiss temperature of most 4$f$ systems is  typically small and thus the onset of spin correlation develops at very low temperature in these quantum materials \cite{Sibille2020,PhysRevLett.115.097202,Gao2019,PhysRevB.105.014441,PhysRevX.1.021002}. Furthermore, the absence of any bifurcation in zero-field cooled (ZFC) and field cooled (FC) susceptibility data (see SI Fig.~2) in the field of 100 Oe suggests that Yb$^{3+}$ spins are not frozen down to 2 K. In order to confirm the absence of spin freezing, we performed  ac susceptibility measurement  down to millikelvin temperatures at different frequencies. Fig.~\ref{st} (d) depicts the real part of ac susceptibility as a function of temperature down to 45 mK and it exhibits a broad maximum around 80 mK which suggests the short range interaction between Yb$^3+$ moments in LYTO \cite{Xing2020,Sarte2021,doi:10.1073/pnas.1906483116}. Importantly, the absence of any frequency dependency of ac susceptibility strongly rules out glassy behavior of Yb$^{3+}$ spins in LYTO. 
The magnetization isotherms at a few selected temperatures are shown in SI Fig.~3 and the  absence of any visible hysteresis  rules out the presence of any ferromagnetic signal.   The calculated  Land\'e $g$ factor, $g$ = 3.46 $\pm$ 0.02,  from the low-temperature Curie-Weiss fit of magnetic susceptibility, is close to that obtained  Land\'e $g$ factor from Brillouin function fit of the magnetization isotherm. A similar value of powder average of Land\'e $g$ factor is also observed in Yb based 3D pyrochlore system suggesting the  single-ion properties of LYTO are expected to  be similar to the pyrochlore system \cite{Sarte2021}.  
\begin{figure*}
	\centering
	\includegraphics[width=\textwidth]{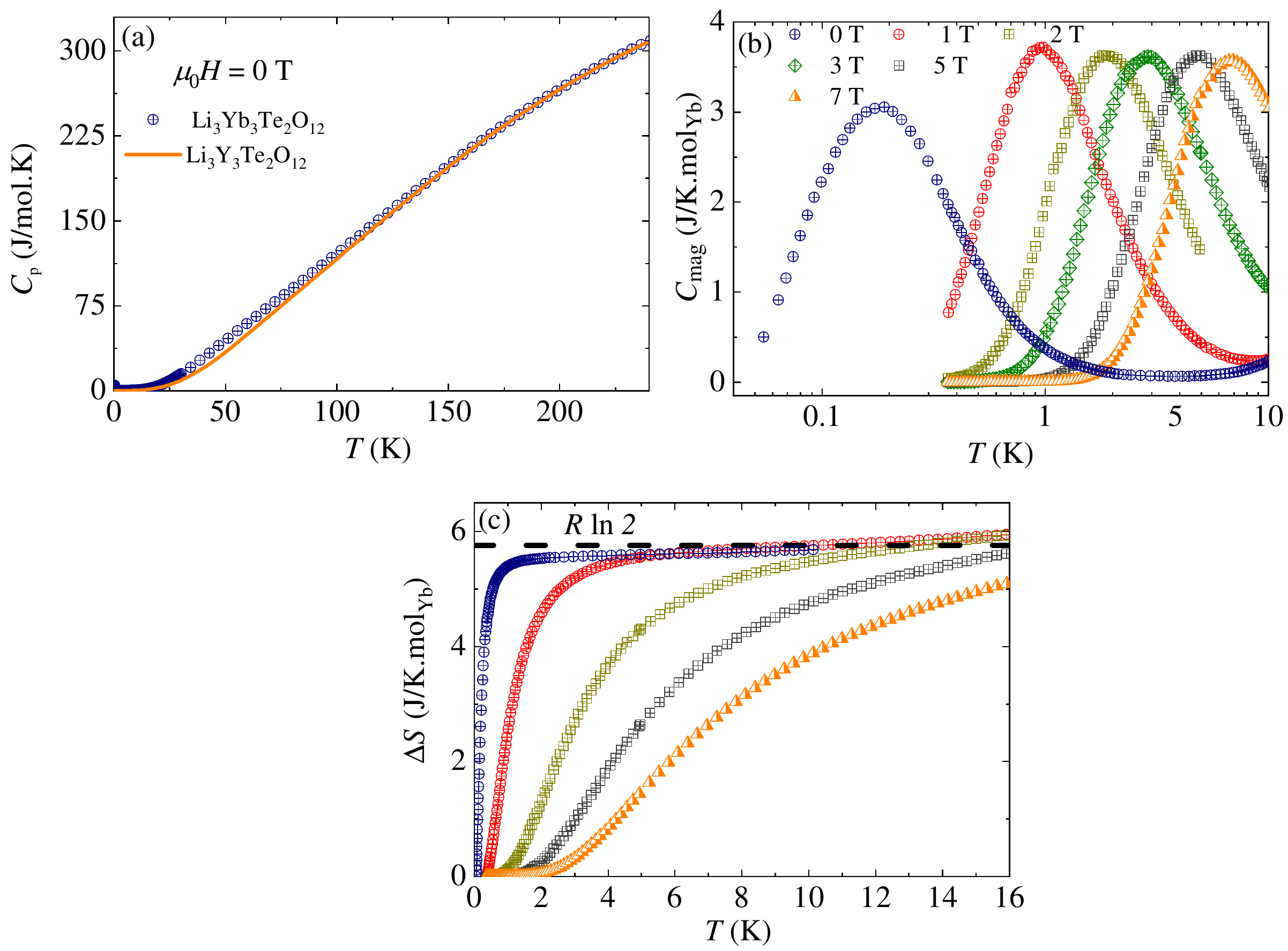}	
	\caption{Specific heat of LYTO in various magnetic fields. (a) The temperature dependence of specific heat ($C_{\rm p}$) of  LYTO and its non-magnetic analog Li$_{3}$Y$_{3}$Te$_{2}$O$_{12}$ in zero-magnetic field. (b) Low-temperature magnetic specific heat ($C_{\rm mag}(T)$)  after subtraction of  lattice  and nuclear  contributions in zero-field as well as non-zero field.
		(c) The temperature dependence of entropy change ($\Delta$S = $\int C_{\rm mag} (T)/T$ d$T$, the temperatures of the color plots correspond to those in (b).)  for magnetic fields up to 7 T with the expected entropy (\textit{R} ln 2 for spin-1/2)  indicated by a black  dashed line. }{\label{heat}}
\end{figure*}
\subsection{Specific heat}
\begin{figure*}
	\centering
	\includegraphics[width= 15 cm, height= 14cm]{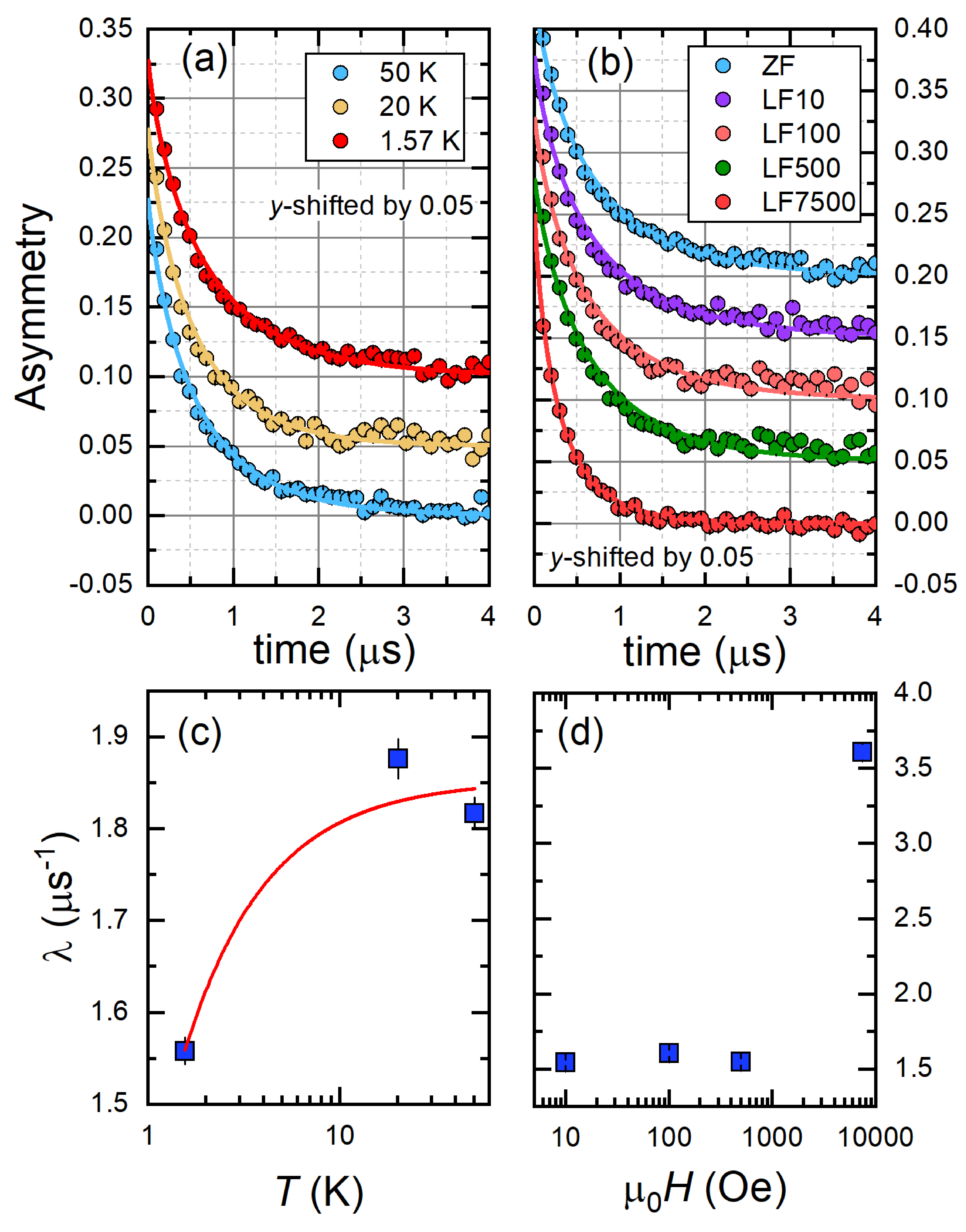}
	\caption{\label{muSR}$\mu$SR data confirming a dynamic ground state  in LYTO. Muon asymmetry decay obtained in ZF (a)  and various LFs at 1.57 K (b) (symbols). The datasets have been shifted vertically by 0.05 for clarity. Lines are fits to a stretched exponential function (see text). The temperature (c) and longitudinal field (d) dependence of the muon spin relaxation rate $\lambda$ extracted respectively from (a) and (b). The red line shows the $1/T$-expansion of $\lambda(T)$ and indicates the presence of antiferromagnetic correlation (see text). Error bars depict an uncertainty of one standard deviation.}
\end{figure*}
Fig.~\ref{heat} (a) shows the total specific heat  measured in zero-magnetic field.
The absence of any $\lambda$-type peak suggests that there is no  phase transition down to 54 mK.  
After subtracting the lattice contribution using the non-magnetic analog Li$_{3}$Y$_{3}$Te$_{2}$O$_{12}$ and the nuclear contribution associated with the hyperfine splitting of the Yb$^{3+}$ nuclear spin multiplet \cite{https://doi.org/10.1002/pssb.200983062}, the resulting temperature dependence of magnetic specific heat, $C_{\rm mag}(T)$ is shown in Fig.~\ref{heat} (b) in the temperature range 54 mK $\leq$ \textit{T} $\leq$ 10 K.  Below 2 K, the zero-field magnetic specific heat, $C_{\rm mag}(T)$  starts increasing  and shows a broad maximum around 0.18 K, which indicates the  presence of  short-range spin correlations between Yb$^{3+}$ moments \cite{Bordelon2019}. The presence of broad maximum around 0.18 K agrees well with the estimated CW temperature at low temperatures and it is consistent with general expectation for a quantum spin liquid candidate \cite{Sarte2021}. \\      
In finite applied magnetic fields, it is observed that the broad maximum shifts to higher temperatures with increased field
which opens gap owing to Zeeman splitting of the lowest Kramers doublet state. 
It is worth noting that  we have not observed any suppression of $C_{\rm mag}(T)$ in presence of magnetic field thus ruling out the presence of disorder \cite{PhysRevB.98.174404}. Therefore, there is no uncertainties in  Schottky contribution due to the effect of disorder in the extracted $C_{\rm mag}(T)$ data.  It is observed that the gap size increases with the increasing magnetic fields  linearly (see SI Figs.~4 and 9) and the fit yields $g$ = 3.34 $\pm$ 0.01 which is  consistent with that obtained Land\'e $g$-factor from bulk magnetization measurements. 
The magnetic field  induced Zeeman gap  in magnetic specific heat data is  observed  at very low-temperature in several rare-earth based spin-liquid candidates  with relatively small exchange interactions compared to the Zeeman splitting \cite{Gao2019,PhysRevB.98.174404}.\\
In Fig.~\ref{heat} (c), we present the estimated entropy change, $\Delta$S, by integrating $C_{\rm mag}(T)/T$ data in the temperature range 0.054 K $\leq$ \textit{T} $\leq$ 16 K and 0.35 K $\leq$ \textit{T} $\leq$ 16 K for zero-field and  finite magnetic fields, respectively.  The total entropy ($R$ln2) expected for $J_{\textnormal{eff}}$ = 1/2 state  is fully recovered for zero-field as well as for non-zero magnetic fields. This provides the concrete evidence of a Kramers doublet ground state of Yb$^{3+}$ spins in LYTO at low temperature.  
\begin{figure*}
	\centering
	\includegraphics[width=0.9\textwidth]{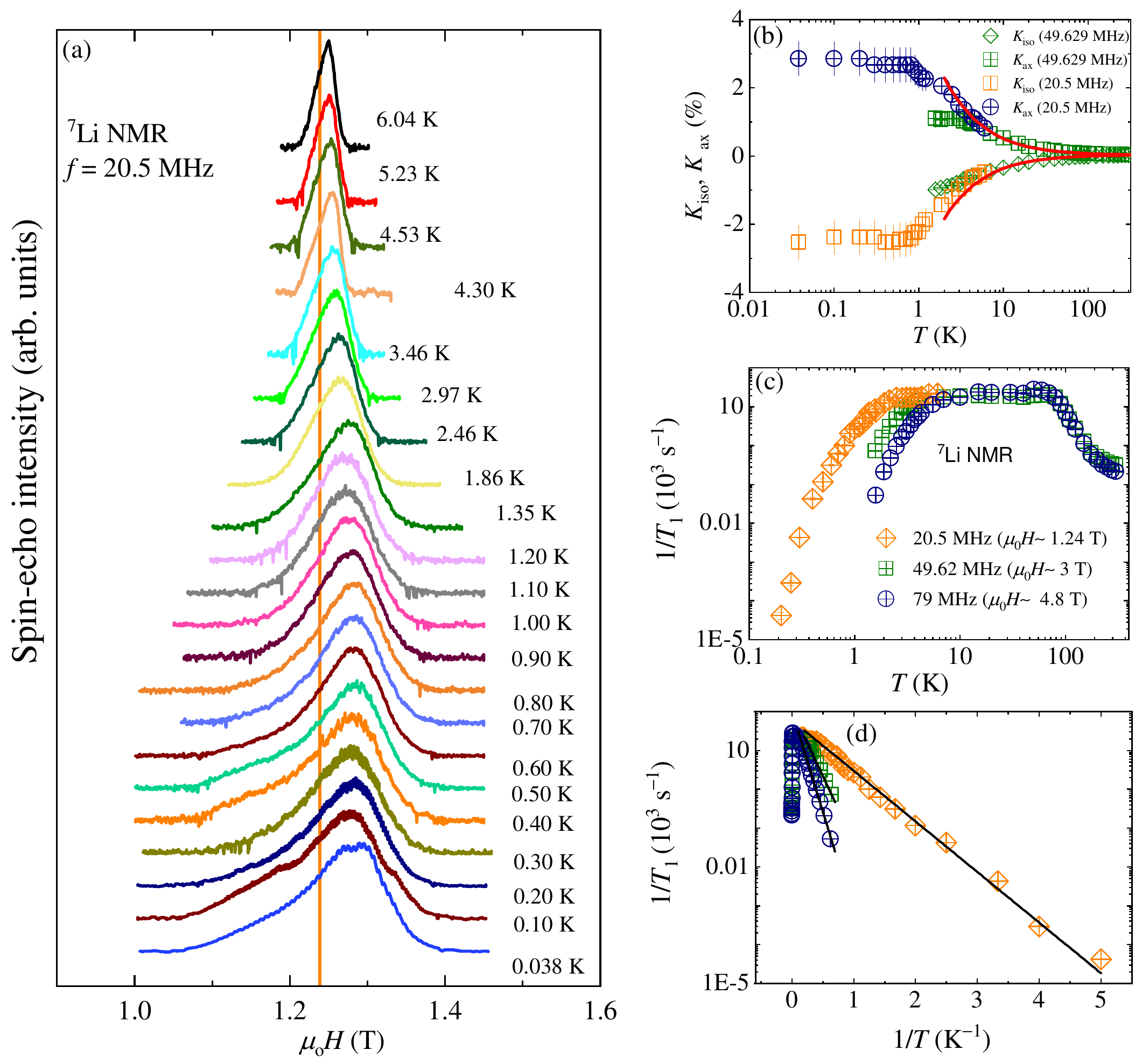}
	\caption{\label{spectra}$^7$Li NMR spectra and  spin-lattice relaxation rate in various magnetic fields reflecting the spin dynamics in the QSL state of LYTO. (a)$^{7}$Li NMR spectra  measured at constant frequency  $\nu$ = 20.5 MHz at a few representative temperatures. The orange vertical line corresponds to zero-shift reference line at 1.238 T. 
		(b)	The temperature dependence of isotropic ($K_{\textnormal{iso}}$) and axial ($K_{\rm ax}$) $^{7}$Li NMR shifts of  LYTO where the solid line is the scaled   magnetic susceptibility measured in  1 T. (c) The temperature dependence of the $^{7}$Li NMR 
		spin-lattice relaxation rate ($T_{\rm 1}^{-\rm 1} $) for three different fields in log-log scale.
		(d) $T_{\rm 1}^{-\rm 1} $ as a function of inverse temperature (\textit{T}$^{-\rm 1}$) for different fields in semi-log scale. The black line represents a fit to a phenomenological model valid for thermally activated behavior of $T_{\rm 1}^{-\rm 1}$  as discussed in the text. Error bars depict an uncertainty of one standard deviation.}
\end{figure*}
\begin{figure*}
	\centering
	\includegraphics[width=\textwidth]{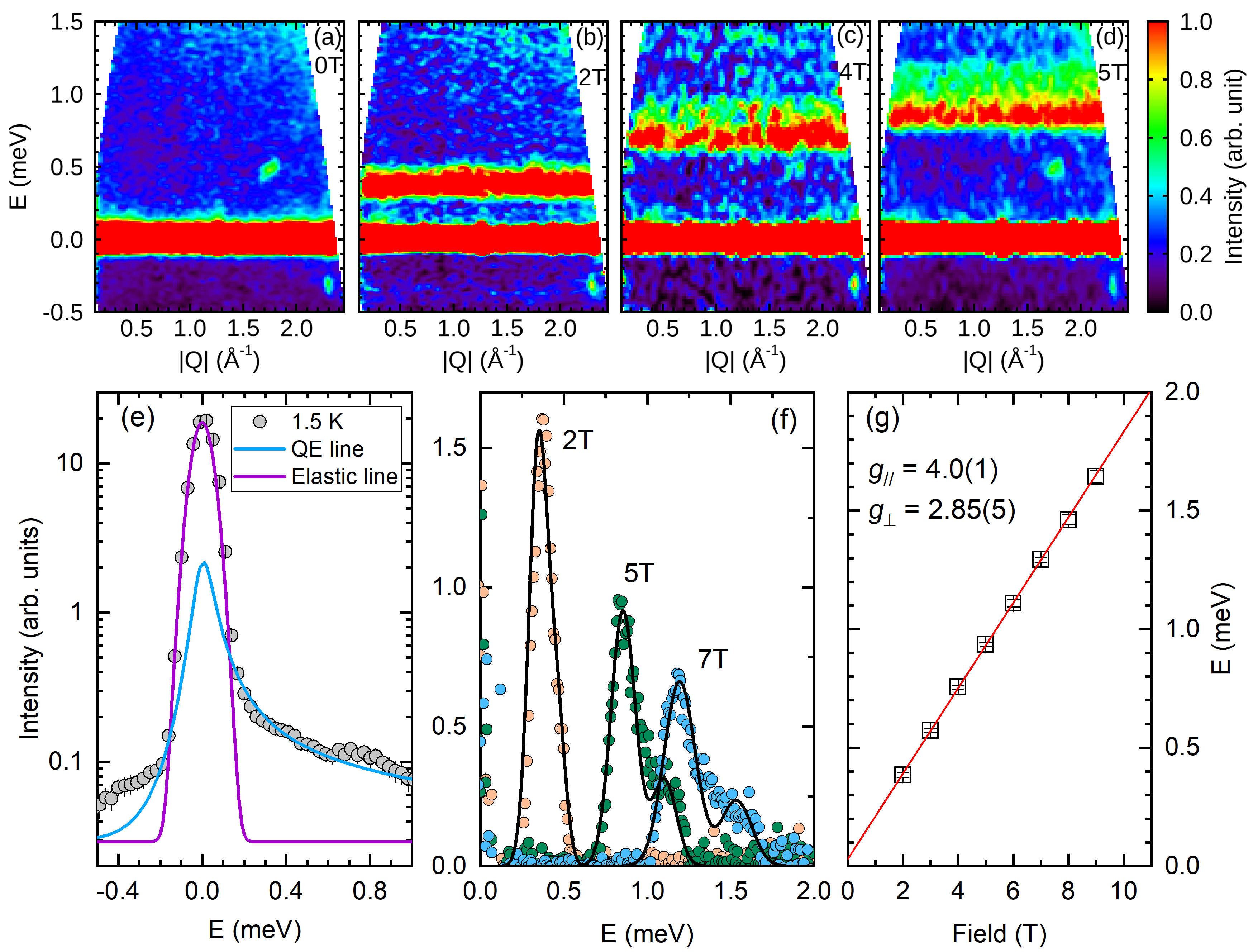}
	\caption{\label{INS}Inelastic neutron scattering. (a-d) Intensity maps of the dynamical structure factor $S(Q,E)$ for polycrystalline LYTO measured at $T=1.5$~K under zero applied field (a), $\mu_0 H=2$~T (b), $\mu_0 H=4$~T (c) and $\mu_0 H=5$~T (d). (e-f) Intensity versus energy for the $Q$-integrated region $[0,1.5]$~{\AA}$^ {-1}$ for the zero-field dataset (e) and some selected datasets under applied fields (from which a field-independent background has been subtracted) (f). Experimental data points are symbols, purple line the estimated elastic contribution of the zero-field data and blue line is the Lorentzian quasi-elastic contribution. Solid black lines in (f) are simulations assuming an axially symmetric $g$-tensor (see text). (g) Median position of the spectrum as a function of the applied field (black squares). Red solid line is a linear fit.}
\end{figure*}
\subsection{Muon spin relaxation} 
We also performed muon spin relaxation ($\mu$SR), to confirm the absence of static magnetism on a local scale. With its spin $I=1/2$ and gyromagnetic ratio $\gamma_{\mu} /2\pi = 135.5$~MHz/T, an implanted muon is an unbeatably sensitive probe of magnetism. Fig.~\ref{muSR} shows the muon asymmetry obtained in zero (ZF) and longitudinal (LF) applied field configurations down to 1.57~K. The ZF relaxation shape is nearly exponential in the $T$-range 1.57-50~K.  A fit of the asymmetry using a stretched exponential function $A(t) = e^{-(\lambda t)^{\beta}}$ leads to large $\beta \sim 0.8-1$ values. This is typical of strongly fluctuating local fields and, indeed, relaxation under LF remains almost unchanged, except for the highest field of 7500 Oe, corroborating  the dynamical character of Yb$^{3+}$ moments. We further investigated this dynamics through temperature dependence of  the muon spin  relaxation rate $\lambda$ (Fig.~\ref{muSR}). Above 20~K, the strong relaxation ($\sim 1.85$~$\mu$s$^{-1}$) is typical of other related $4f$ rare-earth based magnetic materials, such as Yb$_3$Ga$_5$O$_{12}$ or  Yb$_2$Ti$_2$O$_7$, and is consistent with fast fluctuations assuming a similar local field amplitude~\cite{PhysRevLett.91.167201}. The slight decrease of  $\lambda$ at low temperatures can be understood in the context of the presence of antiferromagnetic spin correlations, and the first-order $1/T$-expansion $\lambda (T) = \lambda^{(\infty)}( 1+ \theta_{\rm CW}/T )$ \cite{R_otier_1997} gives a crude estimate of  $\theta_{\rm CW} = -0.25(5)$~K (the red line in Fig.~\ref{muSR} (c)), in agreement with magnetization measurements. 
\subsection{Nuclear magnetic resonance}
We carried out NMR experiments in  LYTO  in order to confirm the absence of magnetic ordering  down to the lowest experimentally accessible temperatures  and to track static and  dynamic susceptibilities. $^{7}$Li NMR  spectra are relatively narrow at room temperature (See SI Fig.~5), however, they exhibit marked broadening below 10 K due to anisotropic interactions and  the development of Yb$^{3+}$ electron spin 
correlations.  We observed no detectable line corresponding to the satellite transitions on both sides of the central transition, which indicates the presence of a very weak quadrupole interaction. Furthermore, the line shape was found  asymmetric indicating the presence of either asymmetry in the hyperfine coupling between the Li nucleus and Yb$^{3+}$ spins or asymmetry in spin susceptibility at low-temperature, both leading to asymmetry in hyperfine fields. 
The  smooth evolution of the  NMR spectra in the entire temperature range rules out the presence of a long-range magnetic ordering at least down to 38 mK. 
The observation of clear shift of spectra in Fig.~\ref{spectra} (a) 
uniquely determines the intrinsic uniform static spin susceptibility on a microscopic scale.  It is  independent and saturates to a finite value in the  low-temperature limit  due to the strong polarization of the Yb$^{3+}$ moments under the applied magnetic field.\\ Fig.~\ref{spectra} (b) presents the temperature dependence of the isotropic NMR shift, $K_{\textnormal{iso}}$, and the  axial shift $K_{\rm ax}$, that were  extracted from the fitting of  NMR spectra. 
At high-temperatures both $K_{\textnormal{iso}}$ and $K_{\rm ax}$ scale with bulk magnetic susceptibility as expected due to significant hyperfine coupling between $^{7}$Li nuclei  and Yb atoms (see SI Fig.~7). The scaling of $K$ with $\chi$  indicates that the macroscopic magnetization is intrinsic and originates from Yb$^{3+}$ electronic moments while any impurity contributions are negligible. The NMR shift  reveals the presence of anisotropic hyperfine field as expected for 4$f$ quantum magnets possibly associated with spin-orbit interactions.\\ 
The   nuclear spin-lattice relaxation rate ($T_{\rm 1}^{- \rm 1}$) probes  the  wave-vector $q$-averaged low-energy spin excitations owing to dynamical electron spin susceptibility in the ground state of correlated quantum materials.  Fig.~\ref{spectra} (c) shows the temperature dependence of $T_{\rm 1}^{-\rm 1}$  in three different magnetic fields, which exhibits the present of three different relaxation regimes.
Upon lowering  the temperature from 300 K,  $T_{\rm 1}^{-\rm 1}$ increases   in the intermediate temperature range and exhibits a plateau around 70 K. Such increase of $T_{\rm 1}^{-\rm 1}$ is not expected for paramagnetic Yb$^{3+}$  spins  but is rather due to crystal electric field effects, where spin fluctuations $\nu$ slow down with decreasing temperature and cause increased NMR relaxation in the fast fluctuation regime, where $1/T_1 \propto 1/\nu$ \cite{PhysRevB.102.045149}.
Below 70 K, $T_{\rm 1}^{-\rm 1}$ remains  constant and field-independent down to 10 K, which suggests that the relaxation is dominated by paramagnetic fluctuations of Yb$^{3+}$ spins.
At low \textit{T}, i.e.,  below $\sim$ 10 K depending on the field, a dramatic field dependence of  $T^{-\rm 1}_{\rm 1}$ develops with a decrease of the $T^{-\rm 1}_{\rm 1}$ value larger than two orders of magnitude in 4.8 T compared to that in 1.24 T at 1.6 K.
$T_{1}^{-1}$   decreases exponentially in the temperature range 0.2 K $\leq$ \textit{T} $\leq$ 10 K, suggesting a gapped excitation spectrum.   It is also observed that, $T_{1}^{-1}$ drops faster in higher magnetic fields which again suggests a field-driven gap as the specific heat data does. 
In Fig.~\ref{spectra} (d), we  present  $T_{\rm 1}^{-\rm 1}$ as a function of inverse temperature ($T^{- \rm 1}$)  in a semi-log plot where the phenomenological model relevant for thermally activated behavior  i.e., $T_{\rm 1}^{- \rm 1}$ $\propto$ exp ($-\Delta/k_{\rm B}T$) yield a straight line with a slope proportional to the gap \cite{PhysRevB.102.045149,PhysRevLett.109.117203}. We observe that the obtained gap  indeed scales linearly with the applied magnetic field (see SI Fig.~9), which is consistent with the
specific heat results. In addition, three distinct energy regimes are also observed in the spin-spin relaxation rate ($T_{\rm 2}^{-\rm 1}$; see SI Fig.~8). The absence of any peak feature in the temperature dependence of $T_{\rm 1}^{- \rm 1}$  rules out the presence of any phase transition  also from the spin dynamics perspective down to 200 mK, the lowest temperature that was reached in our spin-lattice relaxation experiment. 
NMR results thus reveal the presence  of a dynamic ground state  down to 38 mK  with a field-induced gap  owing to Zeeman
splittings of the Kramers doublet state.
\subsection{ Inelastic neutron scattering}
In order to investigate in more detail the nature of  excitations at low-energy, we carried out inelastic neutron scattering measurements on the cold neutron time-of-flight instrument (DCS) at the NCNR (Gaithersburg, MD). Fig.~\ref{INS} shows the dynamical structure factor measured for an incident energy $E_i = 3.27$~meV with a resolution of $0.11$~meV (FWHM). The inelastic spectrum of the zero-field dataset is essentially featureless (Fig.~\ref{INS} (a)). The $Q$-intregated energy cut (Fig.~\ref{INS} (e)) can be well fitted to an elastic Gaussian line plus a quasi-elastic Lorentzian profile, $S(Q,E) \sim (1-e^{-E/ k_{\rm B} T})^{-1} E \Gamma / (E^2 + \Gamma^2)$, accounting for the magnetic contribution. In agreement with the NMR and $\mu$SR results, the Yb$^{3+}$  ions are found to be in a paramagnetic fluctuating regime characterized by a quasi-elastic line with the linewidth $\Gamma = 0.05(2)$~meV setting an upper value of the energy scale of the interactions ($\sim 0.6$~K), which is consistent with all other results. 
Upon increasing the field, a non-dispersive mode emerges from the elastic line (Fig.~\ref{INS} b-d) whose median position depends linearly on the applied field (Fig.~\ref{INS} g). It is ascribed to the Zeeman splitting of the effective spin $J_{\rm eff} $ = 1/2 of the Kramers doublet.
Fig.~\ref{INS} (f) shows $Q$-integrated energy cuts for different applied fields, which further reveal the presence of two humps whose center to center distance increases with fields. This spectrum is characteristic of an axially symmetric $g$-tensor \cite{abragam2012electron}. To further extract the values of the $g$-tensor along the local anisotropy axis, $g_{z} = g_{\parallel}$, and in the perpendicular plane, $g_{xx} = g_{yy} = g_{\perp}$, we simulated the powder-averaged dynamical neutron structure factor (black lines in Fig.~\ref{INS}f) and found $g_{\parallel} = 4.0(1)$ and $g_{\perp} =2.85(5)$, that account for all the datasets from 2 to 9 T. This leads to the powder-average value $g = \sqrt{g_{\parallel} ^2/3+ 2g_{\perp}^2/3} = 3.27(7)$, which is in reasonable agreement with that obtained from other experiments.\\
\section{Summary}
In the 3D
rare-earth based quantum material Li$_{3}$Yb$_{3}$Te$_{2}$O$_{12}$ featuring a perfect hyperkagome lattice constituted by Yb$^{3+}$ ions, our results using several complementary experimental techniques reveal that Yb$^{3+}$ realizes a Kramers doublet and hence   an effective $J_{\rm eff}$ = 1/2 degrees of freedom captures the essence of the  low energy physics  in the ground state.  The Yb$^{3+}$ spins remain dynamic, which is characterized by quasi-elastic INS line in inelastic neutron scattering experiments and $J_{\rm eff}$ = 1/2 moments  interacting with an energy of ~0.6 K, which is rather low but typical for 4$f$-based quantum materials. Furthermore, ac susceptibility data show no sign of spin freezing down to 45 mK. The lack of phase transition down to 54 mK is   evidenced also by the magnetic specific heat data and  a broad maximum around 0.18 K in zero-field is attributed to the presence of short-range spin-correlations. 
The compound thus presents an interesting setting wherein geometrical frustration conspires with quantum fluctuations to evade magnetic long-range ordering as further revealed by our NMR  results down to 38 mK that is consistent with ac susceptibility  and specific heat results, the primary  hallmark of a dynamic ground state.   A  field induced gap was evidenced   in the   specific heat, NMR  relaxation rate and neutron scattering results  which is  reconciled with the  Zeeman splitting of the effective spin $J_{\textnormal{eff}}$ = 1/2 of the Kramers doublet.
The 4$f$ moments on a hyperkagome  lattice- an emblematic frustrated  three dimensional model offer a rare realization of a  spin liquid.  
Our comprehensive results demonstrate a  liquid-like dynamic state in a   rare-earth hyperkagome, which might open the possibility  of stabilizing  spin-orbit driven quantum disordered state in 4$f$ magnets. 
This new family of rare-earth hyperkagome quantum material Li$_{3}$RE$_{3}$Te$_{2}$O$_{12}$ holds great potential to realize spin-orbit driven quantum phenomena and  opens new avenues for modern theoretical approaches in establishing generic  frameworks in the context of spin liquids.	
\section{Acknowledgments} We acknowledge stimulating discussions with P. Mendels. PK acknowledges the funding by the Science
and Engineering Research Board, and Department of Science
and Technology, India through Research Grants.  This research was supported by the U.S. Department of Energy, Office of Basic Energy Sciences, Division of Materials Sciences and Engineering.  Ames Laboratory is operated for the U.S. Department of Energy by Iowa State University under Contract No. DE-AC02-07CH11358. EK and SB acknowledges the France Canada Research Fund for financial support.  AZ acknowledges the financial support of the Slovenian Research agency under program No. P1-0125 and projects No. J1-2461 and No. N1-0148. AMS thanks the URC/FRC of UJ and the SA-NRF for financial support.
\bibliographystyle{apsrev4-1}
\bibliography{LYTO}
\end{document}